\documentclass[journal]{IEEEtran}
\usepackage{amsmath,amsfonts}
\usepackage{algorithmic}
\usepackage{algorithm}
\usepackage{array}
\usepackage[caption=false,font=normalsize,labelfont=sf,textfont=sf]{subfig}
\usepackage{textcomp}
\usepackage{stfloats}
\usepackage{url}
\usepackage{verbatim}
\usepackage{graphicx}
\usepackage{cite}
\begin{document}

\title{Wideband Power Spectrum Sensing: a Fast Practical Solution for Nyquist Folding Receiver}

\author{Kaili Jiang, Dechang Wang, Kailun Tian, Hancong Feng, Yuxin Zhao,\\ Sen Cao, Jian Gao, Xuying Zhang, Yanfei Li, Junyu Yuan, Ying Xiong and Bin Tang
\thanks{Manuscript received August 15, 2023; revised month date, year; accepted month date, year. Date of publication month date, year; date of current version month date, year. This work was supported in part by the Key Project of the National Defense Science and Technology Foundation Strengthening Plan 173 under Grand 2022-JCJQ-ZD-010-12. The associate editor coordinating the review of this manuscript and approving it for publication was xxx. (Corresponding author: Kaili Jiang).}
\thanks{Kaili Jiang, Dechang Wang, Kailun Tian, Hancong Feng, Yuxin Zhao, Junyu Yuan, Ying Xiong and Bin Tang are with the School of Information and Communication Engineering, University of Electronic Science and Technology of China, Chengdu, Sichuan, 611731, China (e-mail: jiangkelly@uestc.edu.cn, c13844033835@163.com, kailun$\_$tian@163.com, 2927282941@qq.com, 1051172535@qq.com, Jyyuan@uestc.edu.cn, yxiong@uestc.edu.cn, bint@uestc.edu.cn).}
\thanks{Sen Cao, Jian Gao, Xuying Zhang and Yanfei Li are with the China Electronics Technology Group Corporation 29th Research Institute, Chengdu, Sichuan, 610036, China (e-mail: caosen@cetc.com.cn, swieegaoj@sina.com, zhxuying@126.com, lyf286lyf@sina.com).}
\thanks{Digital Object Identifier}
}

\markboth{Journal of \LaTeX\ Class Files,~Vol.~14, No.~8, August~2023}%
{Shell \MakeLowercase{\textit{et al.}}: A Sample Article Using IEEEtran.cls for IEEE Journals}


\maketitle

\begin{abstract}
The limited availability of spectrum resources has been growing into a critical problem in wireless communications, remote sensing, and electronic surveillance, etc. To address the high-speed sampling bottleneck of wideband spectrum sensing, a fast and practical solution of power spectrum estimation for Nyquist folding receiver (NYFR) is proposed in this paper. The NYFR architectures is can theoretically achieve the full-band signal sensing with a hundred percent of probability of intercept. But the existing algorithm is difficult to realize in real-time due to its high complexity and complicated calculations. By exploring the sub-sampling principle inherent in NYFR, a computationally efficient method is introduced with compressive covariance sensing. That can be efficient implemented via only the non-uniform fast Fourier transform, fast Fourier transform, and some simple multiplication operations. Meanwhile, the state-of-the-art power spectrum reconstruction model for NYFR of time-domain and frequency-domain is constructed in this paper as a comparison. Furthermore, the computational complexity of the proposed method scales linearly with the Nyquist-rate sampled number of samples and the sparsity of spectrum occupancy. Simulation results and discussion demonstrate that the low complexity in sampling and computation is a more practical solution to meet the real-time wideband spectrum sensing applications.
\end{abstract}


\section{Introduction}
\IEEEPARstart{W}{ith} the rapid application of wireless technology, the spectrum resources are increasingly scarce. There is a growing interest in wideband spectrum sensing for future application in remote sensing, cognitive radios, and Internet of Things \cite{Laake_2022_Remote, Arjoune_2019_Comprehensive, Fang_2021_Recent}, etc. However, the increased demand for wideband spectrum sensing requires a high-speed sampling limited by the Nyquist-Shannon sampling theorem. But the existing analog-to-digital converter (ADC) cannot meet the application requirements of the sampling rate and dynamic range \cite{Manz_2021_Technology}. And the high-speed sampling thus leads to too power-hungry and a prohibitive volume of data for processing, transmitting, and storing. To address these difficulties, many efforts have been made towards the Nyquist sampling-based methods and sub-Nyquist sampling-based methods.

Nyquist sampling-based schemes include direct sampling, sweeping scanner in the time-domain, and channelized receiver in the frequency-domain. With current hardware technologies of direct sampling, the high-speed, high-precision, and wide dynamic range of ADCs are infeasible with low-cost and low-power consumption \cite{Pace_2022_Developing}. The sweeping technique in the time-domain has a low intercept probability for short-lived pulses \cite{Fang_2021_Recent} due to its high latency. And the most widely used currently is the channelized receiver in the frequency-domain, yet which still has some application problems \cite{Tsui_2015_Digital}, i.e. complicated hardware structure, serious RF crosstalk and high power consumption, etc. 

Sub-Nyquist sampling-based schemes are motivated by the compressive sensing (CS) theory \cite{Mishra_2017_Compressivea, Eldar_2012_Compresseda}. The typical compressed sampling architectures contain multi-rate sampling (MRS)\cite{Zhou_2023_Structured, Liu_2022_SparsityBased, Huang_2022_Jointa}, multi-coset sampling (MCS) \cite{Joshi_2020_Learning, Guo_2022_Dualband, Liu_2022_novel}, modulated wideband converter (MWC) \cite{Jiang_2022_Joint, Li_2021_Wideband, Byambadorj_2020_Theoretical}, and Nyquist folding receiver (NYFR) \cite{L3HARRIS__NYFR, Fudge_2022_Multiple, Maleh_2012_Analogtoinformation}, which are utilizing the sparsity structure of the interested inputs. The MRS architecture is currently a preferred shceme for the sparse array signal acquisition, whose performance is also limited by time synchronization accuracy \cite{Huang_2022_Jointa}. The MCS architecture is currently a preferred shceme for the ADC with high-speed and high-precision, which cannot obtain a high significant bit because of the mismatching between multiple channels \cite{Liu_2022_novel}. The MWC architecture depends on the Nyquist-rate pre-randomizing, which is much challenging for the implementation \cite{Byambadorj_2020_Theoretical}. While, this paper focuses on the NYFR architecture, which can theoretically achieve the full-band signal sensing using only a low-speed ADC with the low-speed circuits \cite{Maleh_2012_Analogtoinformation}. 

Although the NYFR can provide a hundred percent probability of intercept (POI) \cite{L3HARRIS__NYFR}, the existing signal processing methods are mainly based on the CS or template matching method. The sparse signal reconstruction via CS-based methods or other optimization methods has high computational and complexity \cite{Wan_2023_Deepa}. And there are high signal-to-noise (SNR) requirements for the information acquisition \cite{Li_2018_Parameterc}. Meanwhile, the inherent bandwidth broadening characteristics of the NYFR outputs could potentially lead to a miss of some weak signals due to the folded noise and harmonics \cite{Tian_2022_Widebanda}.

Such a situation, however, some recent works \cite{Chae_2023_Rethinking, Wang_2018_Phasedarraybased, Ariananda_2012_Compressive} proposed to reconstruct the power spectrum based on the compressive covariance sensing (CCS) mainly for the MWC \cite{Zhang_2021_Joint, Xu_2019_efficient, Cohen_2014_SubNyquist} and MCS \cite{Yang_2020_Fasta, Zhang_2018_Distributed, Ariananda_2011_Multicoset} schemes. Generally, according to the way of computational, the CCS-based wideband power spectrum sensing can be divided into the time-domain approach and the frequency-domain approach. The time-domain approach establishes a relationship between the original inputs and the output samples through the selection matrix with zero and one elements under the equivalent Nyquist-rate sampling \cite{ Yang_2020_Fasta, Xu_2019_efficient, Ariananda_2011_Multicoset}. As well as, the frequency-domain approach builds the relationship between both frequency representations of them \cite{Zhang_2021_Joint, Zhang_2018_Distributed, Cohen_2014_SubNyquist}. 

The CCS-based methods do not need to place any sparsity requirement due to the Toeplitz structure of the covariance matrix. And the second-order statistics can effectively suppress the white noise, which leads to the wideband power spectrum sensing in a low SNR environment. In addition, there is an efficient computation than CS-based methods. Therefore, the CCS-based method is a more practical solution for wideband spectrum sensing. However, the conventional CCS-based methods still have a high computational complexity which cannot meet the real-time wideband power spectrum sensing.

In addition, the power spectrum estimation problem of NYFR has not been discussed in the existing, which can effectively avoid the weak signals to be swamped in the widening bandwidth or folded noise. However, the existing CCS-based methods are not suitable for NYFR, because the NYFR outputs cannot be expressed as a subset of the Nyquist samples. Therefore, a more practical solution for the NYFR is proposed in this paper. By exploring the sub-sampling principle inherent in NYFR, a computationally efficient method is introduced with compressive covariance sensing. That can be efficient implemented via only the non-uniform fast Fourier transform (NUFFT), fast Fourier transform (FFT), inverse fast Fourier transform (IFFT), and some simple multiplication operations. Meanwhile, the state-of-the-art power spectrum reconstruction model for NYFR of time-domain and frequency-domain is constructed in this paper as a comparison.

The rest of the paper is organized as follows. In section \ref{sec2}, the signal model and structure of NYFR is introduced first. Then a fast practical solution of power spectrum estimation for NYFR is proposed in section \ref{sec3}. And section \ref{sec4} analyzes and validates the proposed method through simulation. Then the section \ref{sec5} concludes the paper.

Notations: The lower-case and upper-case bold characters denote the vectors and matrices, respectively. $\mathbb{C}$ and $\mathbb{Z}$ respectively indicate the set of complex values and integer values. $(\cdot)^*$ is the complex conjugation, whereas $(\cdot)^T$ and $(\cdot)^H$ respectively are the transpose and conjugate transpose of a vector or a matrix, and $(\cdot)^{\dag}$ is the inverse or pseudo-inverse operation of a matrix. Then, $\ast$ denotes the convolution operation, $\otimes$ denotes the Kronecker product operation, $\odot$ denotes the Khatri-Rao product operation, and $\circ$ denotes the Hadamard Product operation.  $\text{round}(\cdot)$ represents the rounding operation, and $\text{vec}(\cdot)$ represents the vectorization operation. $\mathcal{FT} \{ \cdot \}$  and $\mathcal{FT}^{-1} \{ \cdot \}$ stands for the Fourier transform (FT) and inverse Fourier transform (IFT) operation, respectively.

\section{Signal Model}
\label{sec2}

\begin{figure}[htp]
	\centering
	\includegraphics[width=8.5cm]{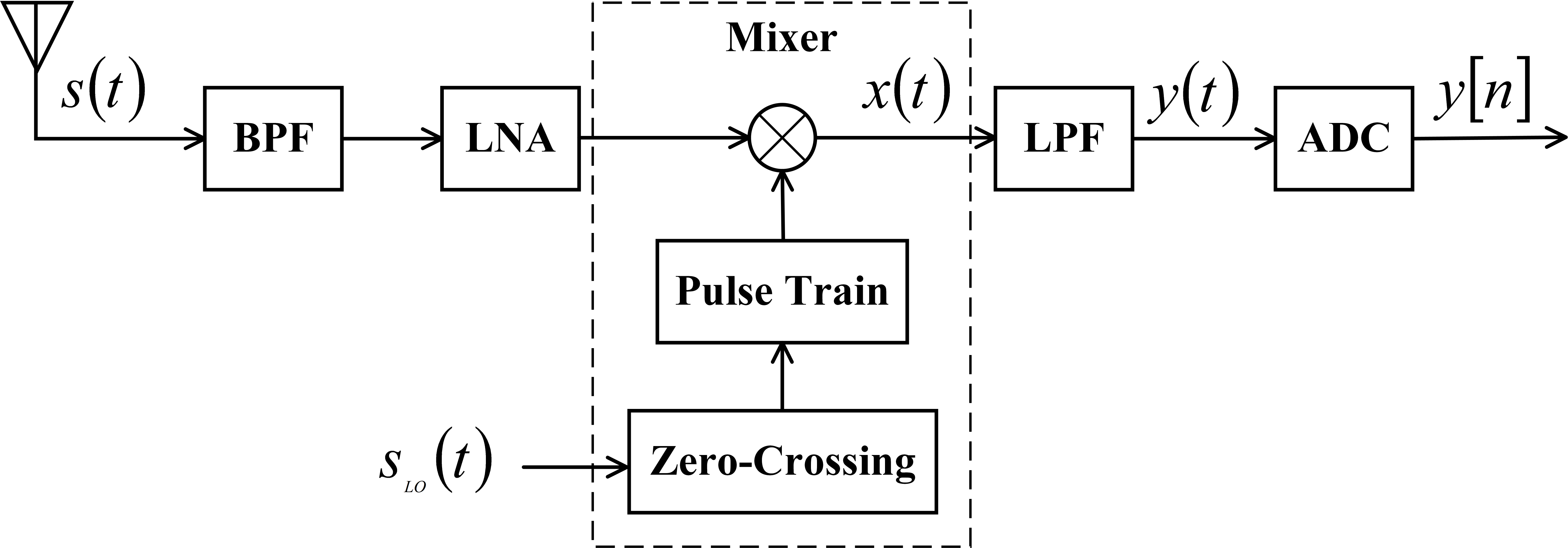}
	\caption{Nyquist folding receiver architecture.}
	\label{NYFR}
\end{figure}

As shown in Figure~\ref{NYFR}, the NYFR is a special secondary sampling structure. The radio frequency (RF) input $s(t)$ is first sampled by the direct RF pulse train without quantization for the 2 to 18GHz band, after passing the preselected band pass filter (BPF) and a low noise amplifier (LNA). And then the baseband input $y(t)$ is obtained from the RF pulse-based sampled signal $x(t)$ at the output of the anti-aliasing low pass filter (LPF). After that the interpolation filter output $y(t)$ is quantized as $y[n]$ by a low-speed ADC as a second sampling. Unlike other CS-based architectures, the NYFR can achieve wideband spectrum sensing at low speed for all circuits avoiding the Nyquist-rate pre-randomization.

The RF pulses are created at each positive zero crossing of the reference local oscillator (LO) signal $s_{LO}(t)$ which is
\begin{equation}
	s_{LO}(t) = sin(\omega_s t + \theta(t))
\end{equation}
where $\omega_s = 2 \pi f_s$ denotes the angular frequency of the first sampling, and $\theta(t)$ is a phase modulation signal. Thus, the instants of positive zero crossing can be given by
\begin{equation}
	\omega_s t + \theta(t) = 2 \pi k, \quad k \in \mathbb{Z}
\end{equation}

According to the scaling properties of dirac sequences, the RF pulse train can be expressed as
\begin{equation}
\begin{split}
	p(t) &= \sum_{k} 2 \pi \delta (\omega_s t + \theta(t) - 2 \pi k) \\
  &\approx \omega_s \sum_{k} e^{jk(\omega_s t + \theta(t))}
\end{split}
\label{eq1}
\end{equation}
and its FT is
\begin{equation}
\begin{split}
  P(\omega) &= \omega_s \sum_{k} \left\{ \delta(\omega - k \omega_s) \ast \mathcal{FT} \{e^{j k \theta(t)}\} \right\} \\ 
  &= \sum_{k} T_k (\omega - k \omega_s)
\end{split}
\end{equation}
where $T_k (\omega) = \omega_s \mathcal{FT} \{e^{j k \theta(t)}\}$. It can be seen that the Fourier spectrum is the periodic extension by period $\omega_s$. However, the spectrum varies within each period, which is related to the period index $k$ and the phase modulation function $\theta(t)$. Thus, the entire spectrum can be distinguished theoretically by this feature. 

\begin{figure}[htp]
	\centering

		\centerline{\includegraphics[width=8.5cm]{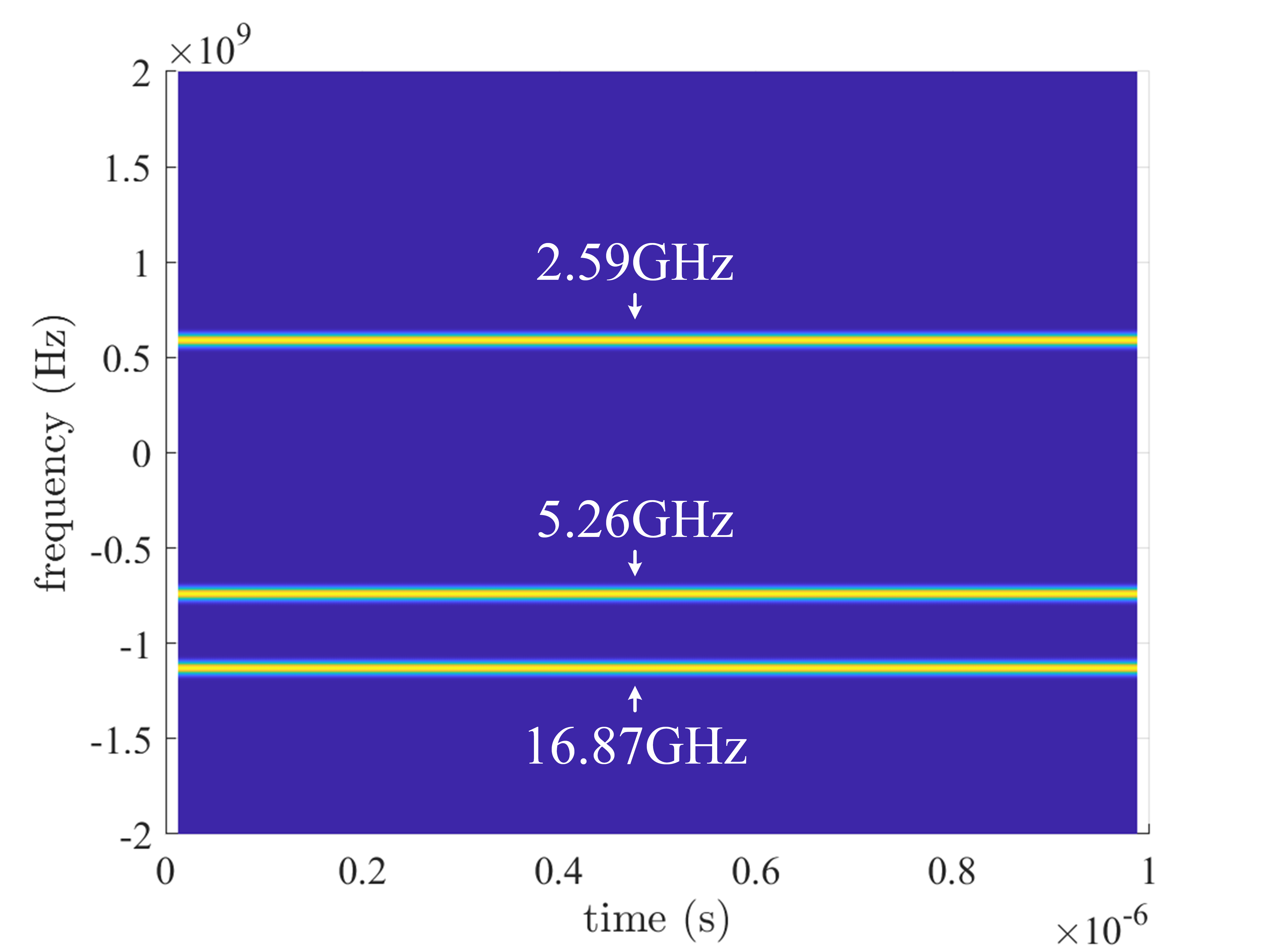}}
		\centerline{(a)}
		\centerline{\centerline{\includegraphics[width=8.5cm]{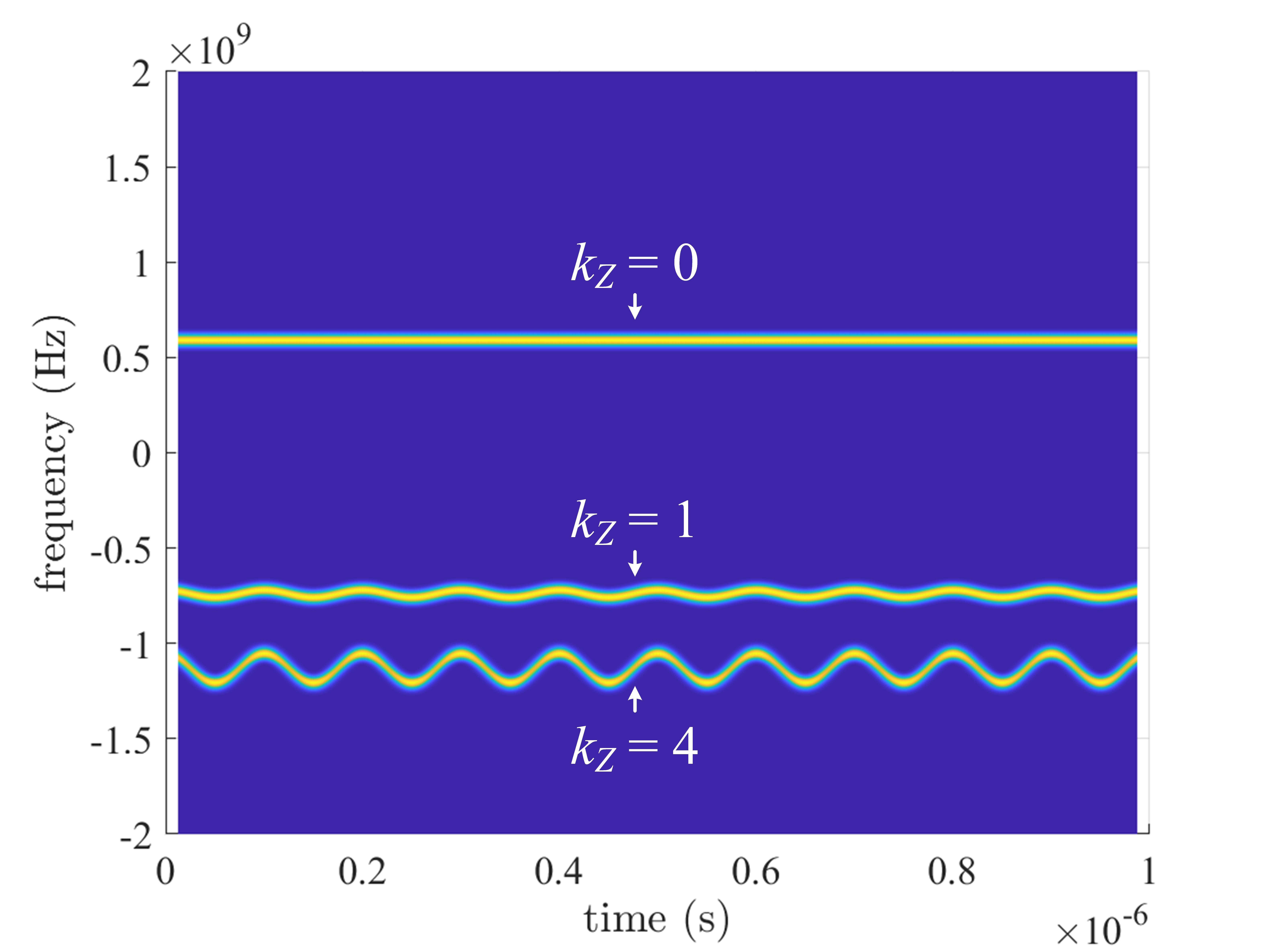}}}
		\centerline{(b)}
	
	\caption{Example NYFR output using noiseless samples collected based $f_s = 4GHz$. (a) Spectrogram of the NYFR output from the uniform subsampling. (b) Spectrogram of the NYFR output from the sinusoidal modulated subsampling.}
	\label{spectrogram}
\end{figure}

Thus, the RF pulse-based sampled signal $x(t)$ can be obtained from 
\begin{equation}
	x(t) = s(t)p(t)
\end{equation}
and its FT can be expressed as
\begin{equation}
\begin{split}
	X(\omega) &= S(\omega) \ast P(\omega) \\
  &= \frac{1}{2 \pi} \sum_{k} \left\{ S(\omega - k \omega_s) \ast T_k (\omega) \right\}
\end{split}
\end{equation}
where $S(\omega)$ is FT of the RF input $s(t)$. At last, the FT of NYFR output $y(t)$ can be obatined from
\begin{equation}
\begin{split}
	Y(\omega) &= X(\omega) H(\omega) \\
  &= \frac{1}{2 \pi} \left[ S(\omega - k_Z \omega_s) \ast T_{k_Z} (\omega) \right]
\end{split}
\end{equation}
where $k_Z$ is the induced Nyquist zone (NZ) index of $s(t)$ with the bandwidth $f_s/2$ for each NZ, which can be calculated from
\begin{equation}
	k_Z = \text{round}({\omega_c}/{\omega_s})
\end{equation}
which decides the sampling rate of ADC and the cutoff frequency of LPF. Therefore, the NYFR output of LPF after can be written as
\begin{equation}
	y(t) = \omega_s s_{k_Z}(t) e^{- j k_Z \theta(t)}
	\label{eq2}
\end{equation}
where $s_{k_Z}(t)$ denotes the down-converted signal from $S(\omega - k_Z \omega_s)$, which can be seen from Figure~\ref{spectrogram}(a). That is an example for the case of the NYFR output in the target spectrum ranges of 2-18GHz, consisting of three inputs at 2.59GHz, 5.26GHz, and 16.87GHz. The original spectrum information is lost in the aliasing from the uniform subsampling. And for the sinusoidally modulated subsampling as shown in Figure~\ref{spectrogram}(b), as expected, different inputs have different scaled versions of the RF sample modulation. That makes original spectrum information and the structure of the signals substantially preserved in the aliasing, unlike many other A2I receivers.

However, the detection performance of the input is limited by bandwidth widening and aliasing. Especially in multi-signal situations, the weak signals are likely to be swamped in the widening bandwidth or the aliasing noise. Meanwhile, since the white noise can be effectively suppressed in the second-order statistics. Therefore, a fast practical solution of wideband power spectrum sensing for the NYFR is proposed in the next section.

\section{Fast power spectrum estimation}
\label{sec3}

Current wideband power spectrum sensing methods can be categorized into the time-domain and frequency-domain power spectrum reconstruction approaches, which are proposed to reconstruct the covariance matrix from the sub-Nyquist sampling samples. Mathematically, the relationship between the original signal and the sub-sampled samples can be given as 
\begin{equation}
	\mathbf{y}[m] = \mathbf{A} \mathbf{s}[n] \qquad \text{or} \qquad \mathbf{y}(\omega_m) = \mathbf{B} \mathbf{s}(\omega)
\end{equation}
where the input vector $\mathbf{s}[n] \in \mathbb{C}^{1 \times N}$ and $\mathbf{s}(\omega) \in \mathbb{C}^{1 \times N}$ denote the Nyquist-rate sampling samples in the time-domain and frequency-domain, respectively. Then, the output vector $\mathbf{y}[m] \in \mathbb{C}^{1 \times M}$ and $\mathbf{y}(\omega_m) \in \mathbb{C}^{1 \times M}$ denote the sub-Nyquist sampling samples in the time-domain and frequency-domain, respectively. In addition, $\mathbf{A} \in \mathbb{C}^{M \times N}$ and $\mathbf{B} \in \mathbb{C}^{M \times N}$ denote the sensing matrix in the time-domain and frequency-domain, respectively. 

This is a universal model for the compressed sampling architectures, and the sensing matrix of NYFR can be constructed followed by \cite{Maleh_2012_Analogtoinformation}, as 
\begin{equation}
\begin{split}
  	\mathbf{A} = & {\left( {\begin{array}{*{20}{c}} 
		{{\mathbf{I}_M}}&{{\mathbf{I}_M}}&\cdots&{{\mathbf{I}_M}}	\end{array}} \right)}	\cdot	\\
		& \left( {\begin{array}{*{20}{c}}
		{{\mathbf{I}_M}}&{}&{}&{}\\
		{}&{{e^{ - j\theta \left( t \right)}}{\mathbf{I}_M}}&{}&{}\\
		{}&{}& \ddots &{}\\
		{}&{}&{}&{{e^{ j {M_Z} \theta \left( t \right)}}{\mathbf{I}_M}}
		\end{array}} \right) \cdot	\\
		& \left( {\begin{array}{*{20}{c}}
		{{\mathbf{\Psi} _M}}&{}&{}&{}\\
		{}&{{\mathbf{\Psi} _M}}&{}&{}\\
		{}&{}& \ddots &{}\\
		{}&{}&{}&{{\mathbf{\Psi} _M}}
		\end{array}} \right) \cdot
		\mathbf{\Psi} _N^{ - 1}
  \end{split}
\end{equation}
where $\mathbf{I}_M$ is an identity matrix of dimension $M$ equals to the number of sub-Nyquist sampling samples, $M_Z$ is the induced modulation index for each NZ, $\mathbf{\Psi}_M$ is the inverse discrete Fourier transform (IDFT) matrix with the rotation factor $\psi = e ^ {j 2 \pi / M}$ and $\mathbf{\Psi} _N^{ - 1}$ is the discrete Fourier transform (DFT) matrix of dimension $N$ equals to the number of Nyquist sampling samples.

As researched in the state-of-arts, the autocorrelation sequence for the time-domain and frequency-domain can be respectively estimated from
\begin{equation}
\begin{split}
	\mathbf{r}_s & = ((\mathbf{A}^{*} \otimes \mathbf{A})\mathbf{C})^{\dag} \text{vec}(\mathbf{R}_y) \qquad \\ & \qquad \qquad \quad \text{or} \\
  \qquad \mathbf{r}_s (\omega) & = (\mathbf{B}^{*} \odot \mathbf{B})^{\dag} \text{vec}(\mathbf{R}_y (\omega)) 
\end{split}
\label{original}
\end{equation}
where the matrix $\mathbf{C} \in \mathbb{C}^{N^2 \times N}$ is a corresponding selection matrix for the sensing system of only zero and one element. And the matrix $(\mathbf{A}^{*} \otimes \mathbf{A})\mathbf{C}$ is of size $M^2 \times N$, which has the same size as the matrix $\mathbf{B}^{*} \odot \mathbf{B}$. Meanwhile, these two matrices can be computed offline in advance. 

To see this, there is an example of wideband spectrum sensing based on the MWC architecture with 1GHz bandwidth, requiring a spectrum resolution of 10kHz. The time-domain approach involves at least $10^{14}$ floating-point operations in total, assuming that the number of sampling branches is set as 8, the downsampling factor sets as 25, then the number of output samples need to be 4000, and the number of samples set as 100 used to calculate the correlation matrix. And the more efficient time-domain approach has the same computational complexity as the frequency-domain approach, which involves more than $10^{7}$ floating-point operations in total for the same assumption. However, that is unsuitable for real-time applications with a high compression ratio. 

To address this issue, a fast practical solution with computationally efficient is proposed for NYFR. Considering, the RF pulse-based sampled vector $\mathbf{x}[n] \in \mathbb{C}^{N \times 1}$ under the Nyquist-rate sampled can be obtained from 
\begin{equation}
	\mathbf{x}[n] = \mathbf{s}[n] \circ \mathbf{p}[n]
\end{equation}
where the RF input vector $\mathbf{s}[n] \in \mathbb{C}^{N \times 1}$ and the RF pulse train vector $\mathbf{p}[n] \in \mathbb{C}^{N \times 1}$ are all obtained by sampling the analog signal with a Nyquist sampling rate. Therefore, the autocorrelation sequence elements $r_x [k]$ can be estimated by
\begin{equation}
	\begin{aligned}
		r_x [k] 
		&= \frac{1}{N} \sum_{k=0}^{N-1} x[n] x^{*}[n-k]\\
		&= \frac{1}{N} \cdot \mathbf{x}[n] \mathbf{x}^H [n-k]\\
		&= \frac{1}{N} \cdot (\mathbf{s}[n] \circ \mathbf{p}[n]) (\mathbf{s}^H [n-k] \circ \mathbf{p}^H [n-k])\\
		&= \frac{1}{N} \cdot (\mathbf{s}[n] \mathbf{s}^H [n-k]) \circ (\mathbf{p}[n] \mathbf{p}^H [n-k])\\
		&= r_s [k] \circ r_p [k]
	\end{aligned}
\end{equation}
where $\lvert k \rvert \leq N-1$. Thus, the power spectrum can be obtained by the FT of the autocorrelation sequence $ \{ r_s [k] \}$ through obtaining the autocorrelation sequence $\{ r_x [k] \}$ and $\{ r_p [k] \}$. 

For the RF pulse train, a widely-used estimation of its autocorrelation sequence is given as
\begin{equation}
	r_p [k] = \frac{1}{N} \sum_{k=0}^{N-1} p[n] p^{*}[n-k], \quad \lvert k \rvert \leq N-1
	\label{eq15}
\end{equation}
where $p[n]$ can be constructed from the analog signal as (\ref{eq1}) with a Nyquist sampling rate. And that can be computed offline in advance. 

For the RF pulse-based sampled signal, the harmonic components in only one NZ are selected as the NYFR output through LPF. But it's worth noting that the harmonics for each NZ are a scaled version of the same LO modulation function, where the scale factor is related to the NZ index. Whereas, the introduction of the LO modulation function is brought by the non-uniform sampling moments. As a result, the FT spectrum of $x(t)$ with the Nyquist sampling rate can be reconstructed by the non-uniform Fourier transform (NFT) \cite{Wei_2022_Nonuniform}, as follows  
\begin{equation}
	Y' (\omega) = \sum_{m=0}^{M-1} y(t_m) e^{- j \omega t'_m} 
\end{equation}
where $y(t_m)$ is the NYFR output as (\ref{eq2}) from the low-speed ADC, i.e. 
\begin{equation}
	Y' (\omega) 
		= \omega_s \sum_{m=0}^{M-1} s_{k_Z}(t_m) \cdot  e^{- j k_Z \theta(t_m)} \cdot e^{- j \omega t'_m}
	\label{eq4}
\end{equation}
where $\omega \in [ - \omega_s K_Z /2 , \omega_s K_Z /2 ]$ with the total number $K_Z$ of NZ index covered by the receiver, and $t_m$ is the uniform sampling instants. While, $t'_m$ is the non-uniform sampling instants, which is given by 
\begin{equation}
	t'_m = \frac{2 \pi m - \theta(t_m)}{\omega_s}, \quad m \in \mathbb{Z}
	\label{eq3}
\end{equation}	
where the degree of non-uniformity is determined by $\theta(t)$, which degenerates to the uniform sampling when $\theta(t) = 0$.

Substituting (\ref{eq3}) into (\ref{eq4}) yields
\begin{equation}
	Y' (\omega) = \omega_s \sum_{m=0}^{M-1} s_{k_Z}(t_m) \cdot  e^{- j k_Z \theta(t_m)} \cdot e^{- j \omega \frac{2 \pi m - \theta(t_m)}{\omega_s}}
\end{equation}
and let $\omega = l \omega_s + \omega^{'}$, we have
\begin{equation}
	\begin{aligned}
		Y' (l \omega_s + \omega')
			&= \omega_s \sum_{m=0}^{M-1} s_{k_Z}(t_m) \cdot  e^{- j k_Z \theta(t_m)} \\
      & \quad \cdot e^{- j (l \omega_s + \omega') \frac{2 \pi m - \theta(t_m)}{\omega_s}}\\
			&= \omega_s \sum_{m=0}^{M-1} s_{k_Z}(t_m) \cdot  e^{- j k_Z \theta(t_m)} \\
      & \quad \cdot e^{- j l \omega_s \frac{2 \pi m - \theta(t_m)}{\omega_s}} \cdot e^{- j \omega' \frac{2 \pi m - \theta(t_m)}{\omega_s}} \\
			&= \omega_s \sum_{m=0}^{M-1} s_{k_Z}(t_m) \cdot  e^{- j k_Z \theta(t_m)} \\
      & \quad \cdot e^{- j 2 \pi l m} \cdot e^{ j l \theta(t_m)} \cdot e^{- j \omega' \frac{2 \pi m - \theta(t_m)}{\omega_s}} \\
			&= \omega_s \sum_{m=0}^{M-1} s_{k_Z}(t_m) \cdot  e^{- j (k_Z-l) \theta(t_m)} \\
      & \quad \cdot e^{- j \omega' \frac{2 \pi m - \theta(t_m)}{\omega_s}} \\
	\end{aligned}
\end{equation}
where $\omega' \in [ - \omega_s /2 , \omega_s /2 ]$ and $l=0,1,\dots,K_Z-1$. Then, bringing (\ref{eq3}) back to the above equation, it is easy to know that
\begin{equation}
	Y' (l \omega_s + \omega') = \omega_s \sum_{m=0}^{M-1} s_{k_Z}(t_m) \cdot  e^{- j (k_Z-l) \theta(t_m)} \cdot e^{- j \omega' t'_m}
\end{equation}

Therefore, the RF pulse-based sampled sequence can be obtained from
\begin{equation}
	\hat{\mathbf{x}}[n] = \mathcal{FT}^{-1} \{ \mathbf{Y}' (\omega) \}
	\label{eq22}
\end{equation}
and the estimation of its autocorrelation sequence is given as
\begin{equation}
	r_x [k] = \frac{1}{N} \sum_{n=0}^{N-1} \hat{x}[n]\hat{x}^{*}[n-k], \quad \lvert k \rvert \leq N-1
	\label{eq23}
\end{equation}

After obtaining the sequences $\{ r_x [k] \}$ and $\{ r_p [k] \}$, the autocorrelation sequence of the RF input can be computed via
\begin{equation}
	r_s [k] = r_x [k] ./ r_p [k]
	\label{eq24}
\end{equation}

\begin{figure}[htp]
	\centering
	\includegraphics[width=6cm]{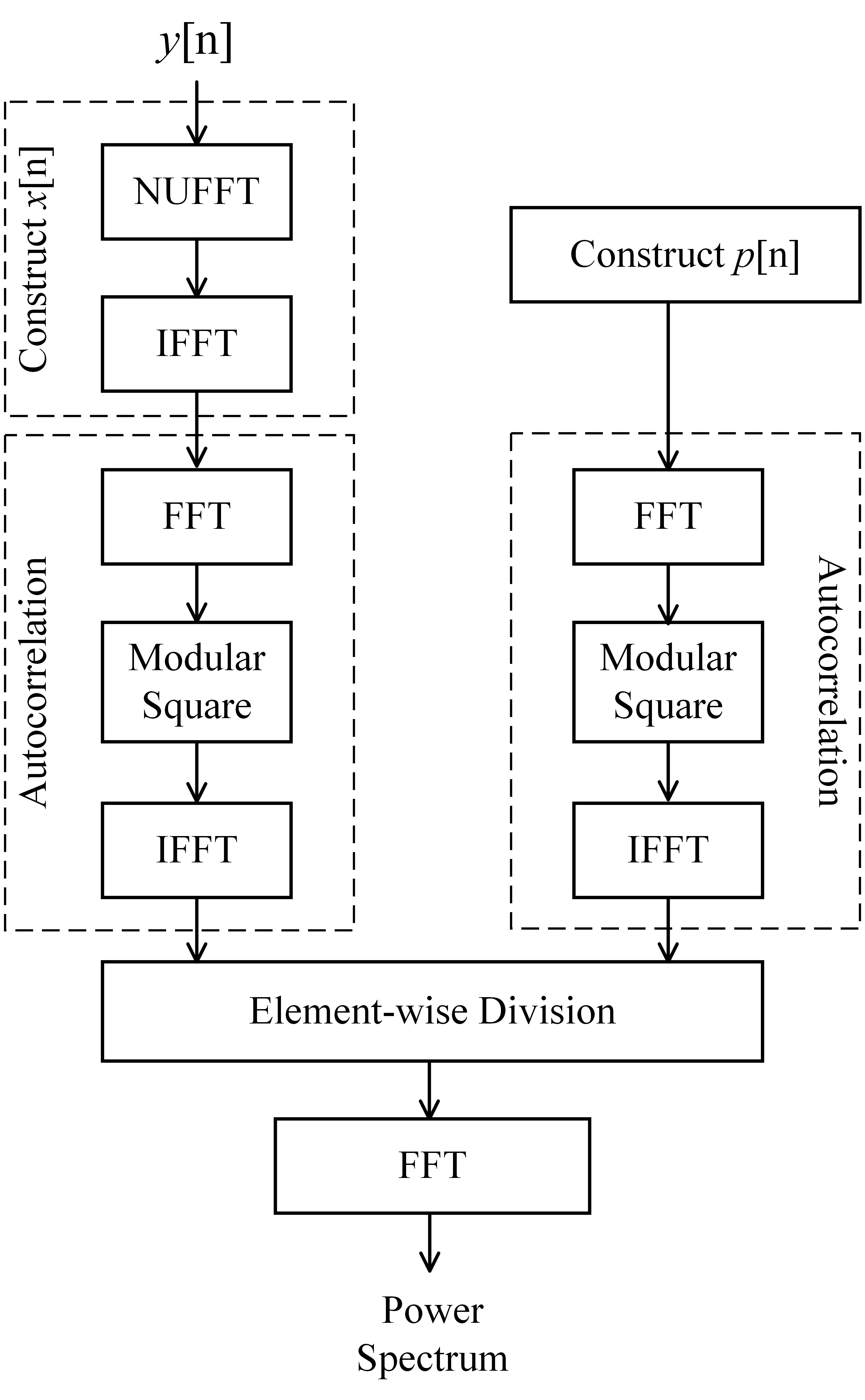}
	\caption{Block diagram of the proposed wideband power spectrum sensing method for NYFR.}
	\label{method}
\end{figure}

The block diagram of the fast practical solution for NYFR is shown in Figure~\ref{method}, where the autocorrelation calculation can be realized by FT. Then the proposed method only involves FFT, IFFT, NUFFT, and some simple multiplication operations. That is a more suitable solution to meet the real-time wideband power spectrum sensing.

\section{Simulation and discussion}
\label{sec4}

In this section, the performance of the proposed wideband power spectrum sensing method for NYFR is discussed through simulation. In the experiments, the power spectrum sensing coverage is considered in the target spectrum ranges of 2-18GHz. The instantaneous bandwidth for NYFR is 16GHz, which is limited by the BPF. Then the sampling rate of ADC is set to 4GHz, which decides the 2GHz bandwidth of the LPF. Meanwhile, the NYFR system covers 8 NZs with a bandwidth of 2GHz for each one. While, assuming that the phase modulation signal of LO is sinusoidal modulation, with the amplitude and frequency are 2 and 20MHz, respectively. And there are 4000 measurements and 32000 Nyquist samples collected within 1ms duration. 

In addition, the CCS-based methods are suitable for wide-sense stationary signals and cyclostationary signals. Therefore, the simulations take the mono-frequency pulse (MP) signal, binary phase shift keying (BPSK) signal and linear frequency modulation (LFM) signal as an example. Moreover, the accuracy is adopted to evaluate the performance of the proposed wideband power spectrum sensing method, defined as
\begin{equation}
	\text{Accuracy} = \frac{\text{Number of eligible experiments}}{\text{Total number of experiments}} \times 100\%
\end{equation}
where the total number of experiments is one hundred under the certain experimental condition. And there is one hundred Monte Carlo trials in the simulation.

\begin{figure}[htp]
	\includegraphics[width=8.5cm]{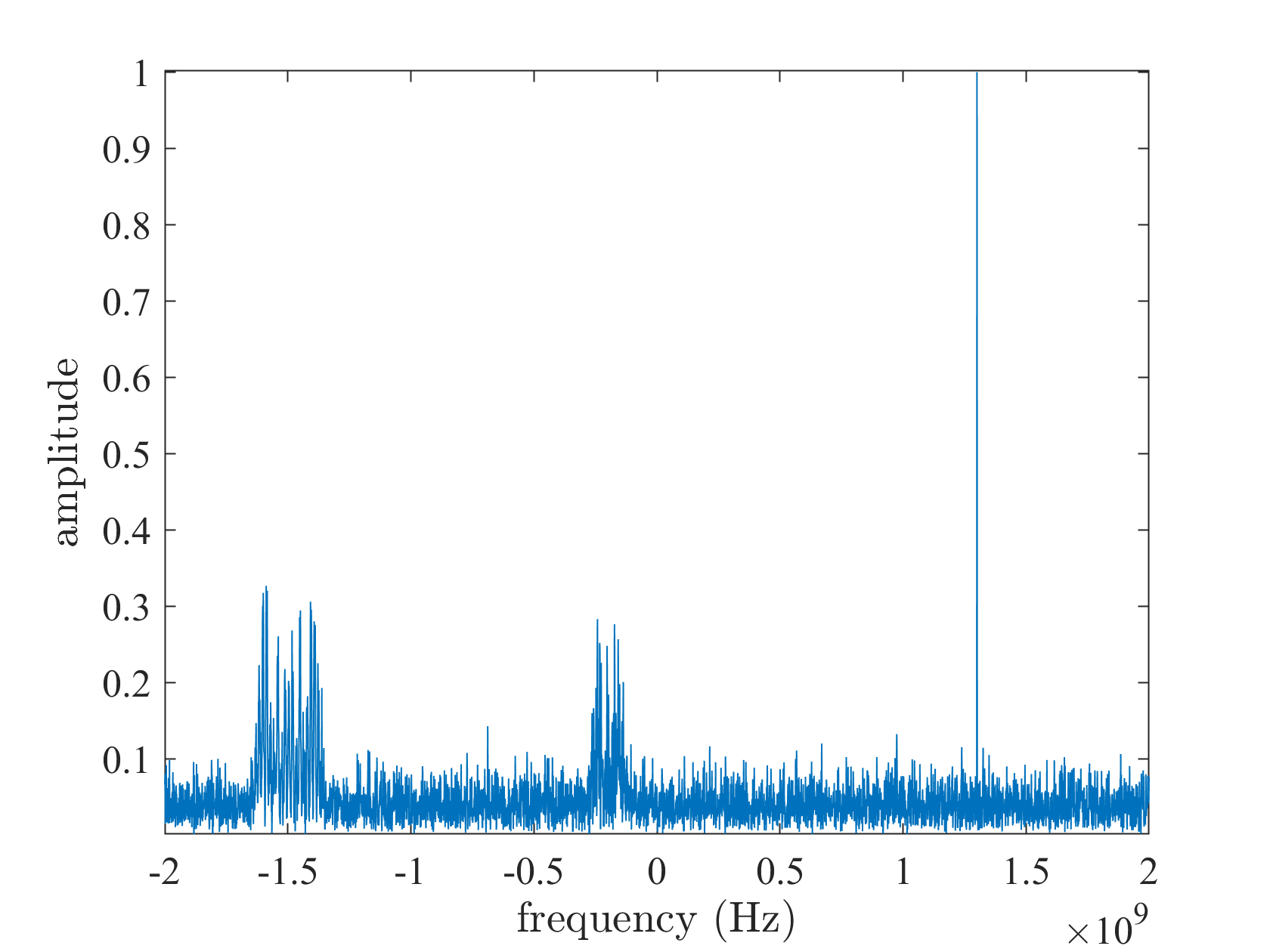}
	\caption{Frequency spectrum of the NYFR output from LPF (SNR = 10).}
	\label{CCS_yfft}
\end{figure}

\begin{figure}[htp]
	\includegraphics[width=8.5cm]{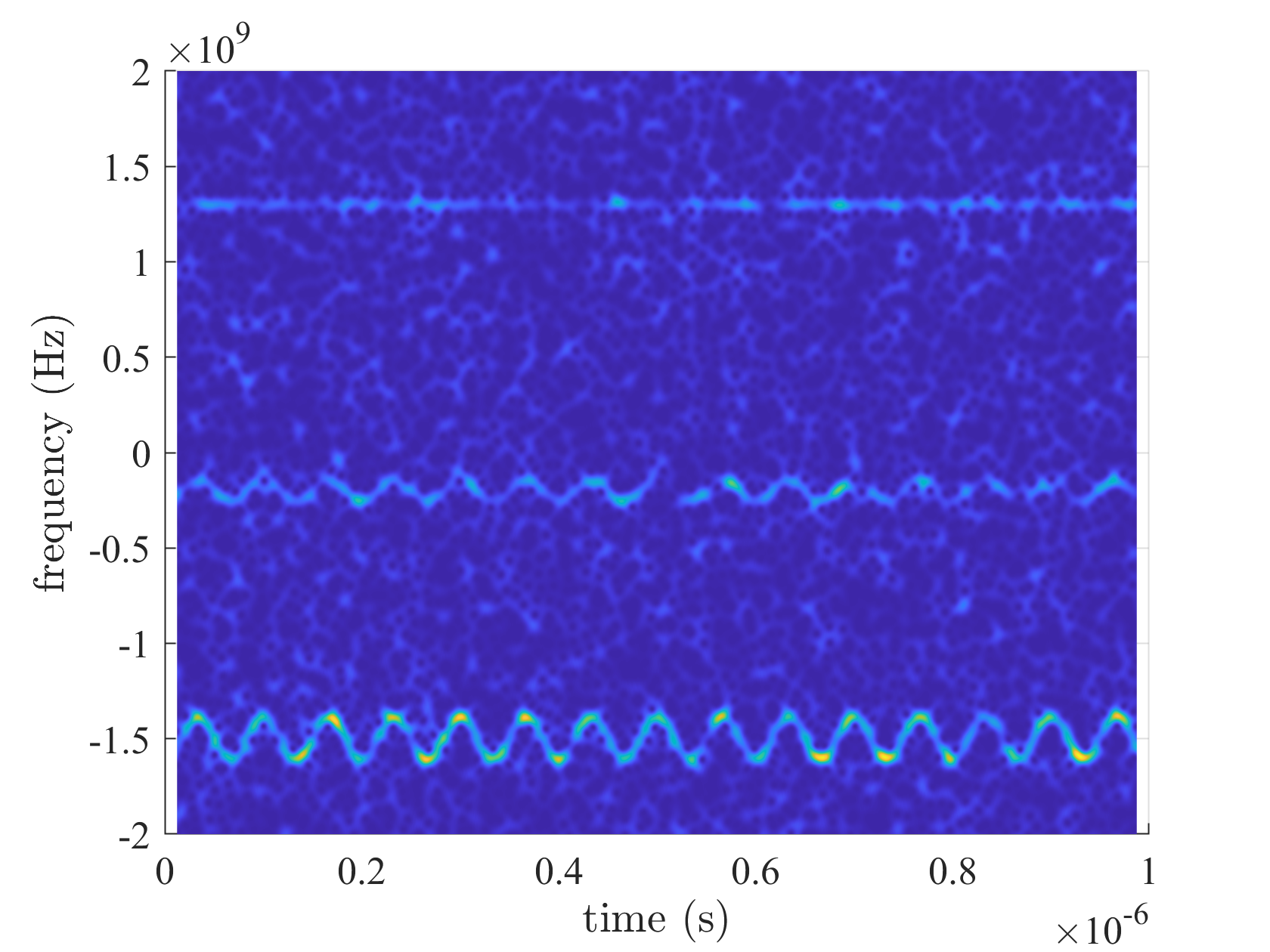}
	\caption{Spectrogram of the NYFR output from LPF (SNR = 10).}
	\label{CCS_TF}
\end{figure}

\begin{figure}[htp]
	\includegraphics[width=8.5cm]{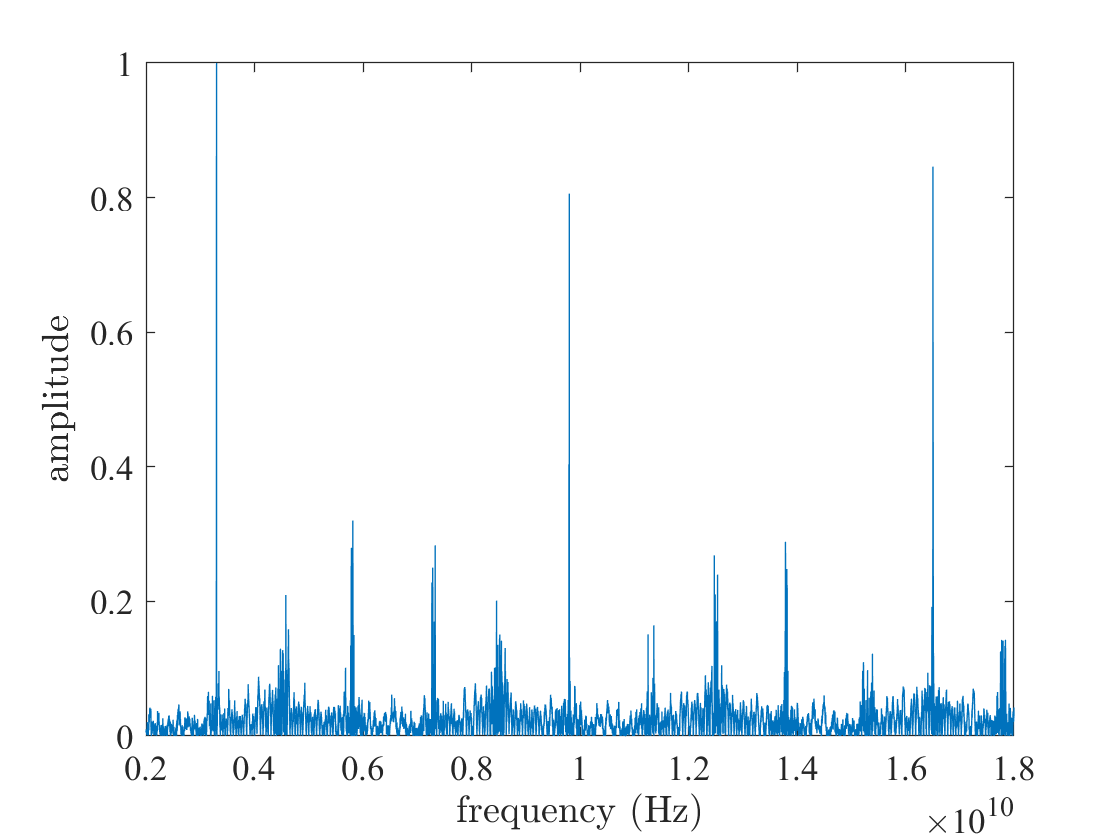}
	\caption{Reconstructed power spectrum of the NYFR output (SNR = 10).}
	\label{CCS}
\end{figure}

Herein, the result on the Fourier spectrum, short-time Fourier spectrogram, and proposed wideband power spectrum reconstruction are first displayed in Figure~\ref{CCS_yfft}, Figure~\ref{CCS_TF} and Figure~\ref{CCS}, respectively. The example includes three signals of MP, BPSK, and LFM, whose carrier frequencies are set to 1.3, 7.8, and 14.5 GHz respectively. There are the same amplitude and initial phase for the inputs. The symbols of the BPSK signal are set to '1001100110' and the bandwidth of the LFM signal is set to 8MHz. Meanwhile, the signals cover the entire observation duration with an input SNR sets to 10dB. As the earlier simulation settings, the RF non-uniform sampling clock is a jittered sinusoidal modulation centered at 4GHz and varying approximately by 40MHz over one period. 

As shown in Figure~\ref{CCS_yfft}, the original spectrum information of inputs is downconverted to 1.3, -0.2, and -1.5GHz, based on the down-conversion factor as 4GHz. And the expected NZ index for the three signals is 0, 2, and 4, respectively. Therefore, the bandwidth spread of the LFM signal is the largest. As expected, the induced modulation of NYFR output is a scaled version of the RF non-uniform sampling clock, which is shown in Figure~\ref{CCS_TF}. The resolution of the spectrogram is insufficient, however, because of the SNR loss from sub-Nyquist sampling. In this example, there is a 12dB SNR loss based on four times folding, leading to an output SNR of NYFR as -2dB. To simplify, the assumptions in this section are all RF input SNR. 

Furthermore, the reconstructed power spectrum of the NYFR output for the entire frequency range is shown in Figure~\ref{CCS}. It is visible that all input frequencies can be identified correctly. While, the power of the pseudo spectrum is relatively higher, which is related to the original spectrum information of inputs and the system parameter selection of NYFR. The pseudo spectrum can be suppressed by filtering for each NZ and system optimization design, but that's beside the point here.

\begin{figure}[htp]
	\includegraphics[width=8.5cm]{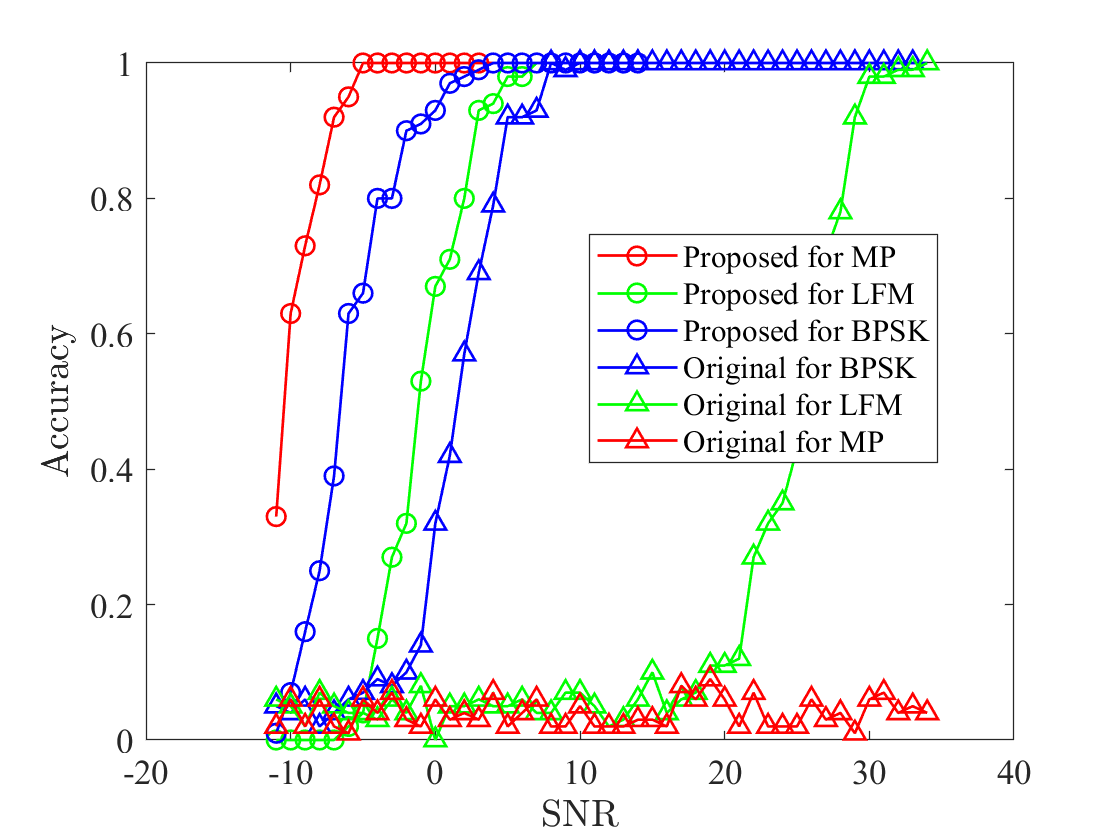}
	\caption{Power spectrum sensing performance versus SNR (K = 1).}
	\label{SNR}
\end{figure}

In Figure~\ref{SNR}, accuracy results are compared as a function of the input SNR, where $K = 1$ is assumed and the carrier frequency is randomly selected in the 2-18GHz. The symbol rate of the BPSK signal is set to 10M symbols per second with random code. And the bandwidth of the LFM signal is set to 10MHz. Meanwhile, the pulse length is fixed at 500ns with a start time randomly distributed throughout the observation duration. 

It can be seen that the accuracy result of the proposed method for the MP signal tends to be stabilized when SNR is greater than -5dB. In contrast, the original method can not accurately identify the MP signal, which is due to the sensing model mismatching under the noise disturbance. While as for the BPSK signal, the accuracy result of the proposed method and original method tend to be stabilized when SNR is greater than 4dB and 8dB, respectively, yields that the power spectrum of cyclostationary signals can be estimated accurately. Then for the LFM signal, the accuracy result of the proposed method tends to be stabilized when SNR is greater than 8dB, which is much more than the original method with 33dB. 

\begin{figure}[htp]
	\includegraphics[width=8.5cm]{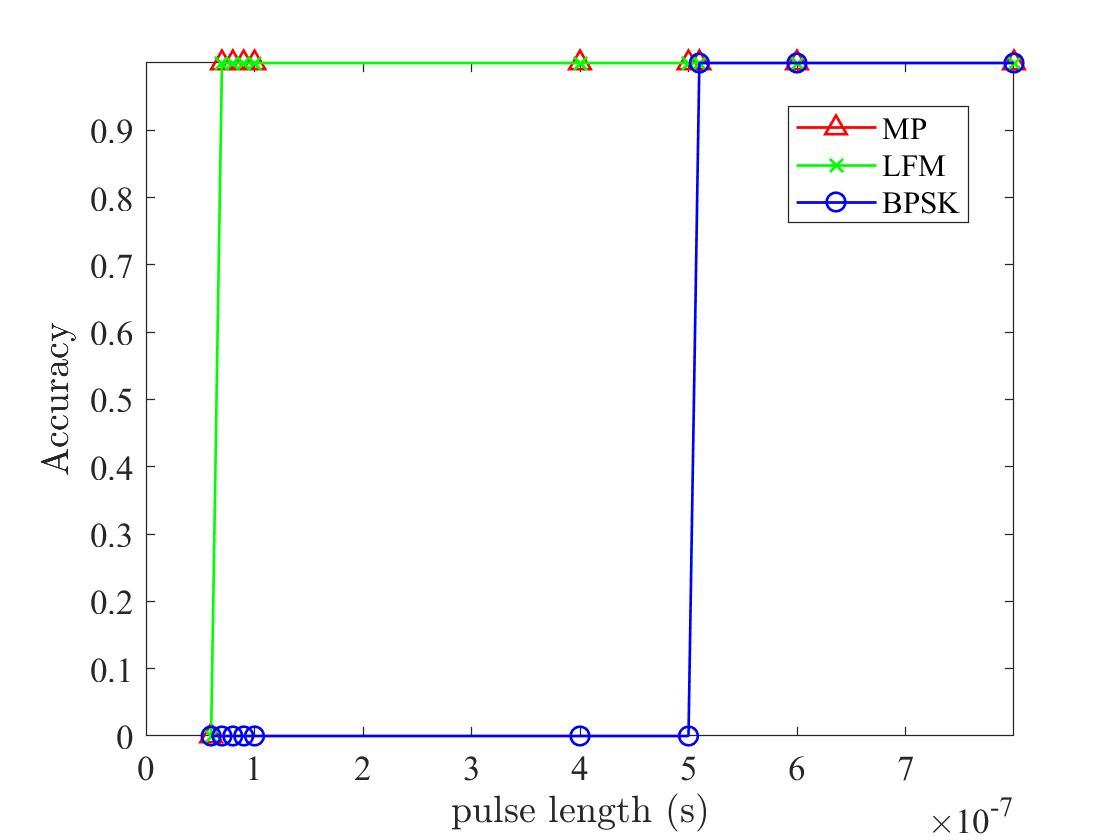}
	\caption{Pulse length requirement for the proposed power spectrum sensing (K = 1, SNR = -5dB).}
	\label{pulselength}
\end{figure}

The pulse length requirement of inputs is then considered for the proposed power spectrum sensing method, whose sensing performance is of particular interest in radar and communication systems. As shown in Figure~\ref{pulselength}, there is $K = 1$ carrier frequency which is selected in the 2-18GHz randomly, with an input SNR sets to -5dB. Moreover, the start time of the pulses is randomly distributed throughout the observation duration. In addition, the symbol rate of the BPSK signal and the bandwidth of the LFM signal are both set to 10MHz. As a result, the pulse length requirement of the MP signal and the BPSK signal is at least 70ns, which need to be greater than 500ns for the LFM signal. 

\begin{figure}[htp]
	\includegraphics[width=8.5cm]{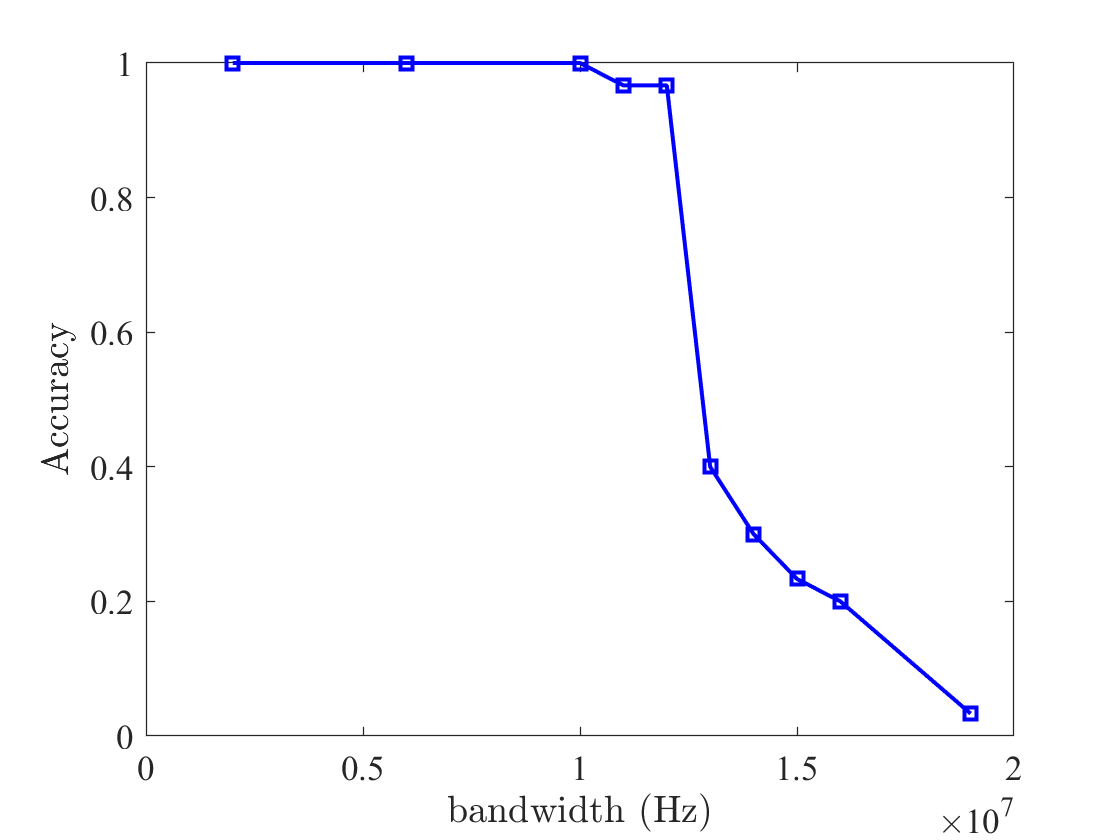}
	\caption{Power spectrum sensing performance versus bandwidth of LFM signal (K = 1, SNR = -5dB).}
	\label{bandwidth}
\end{figure}

Furthermore, the power spectrum sensing performance versus the bandwidth of the LFM signal is depicted in Figure~\ref{bandwidth}, where the pulse length is set to 500ns. It is observed that the performance deteriorates when the bandwidth exceeds 10MHz, which is affected by the phase modulation signal of LO. The amplitude, frequency, and even the symbol sequence of LO all determine the influence of system performance.

\begin{figure}[htp]
	\includegraphics[width=8.5cm]{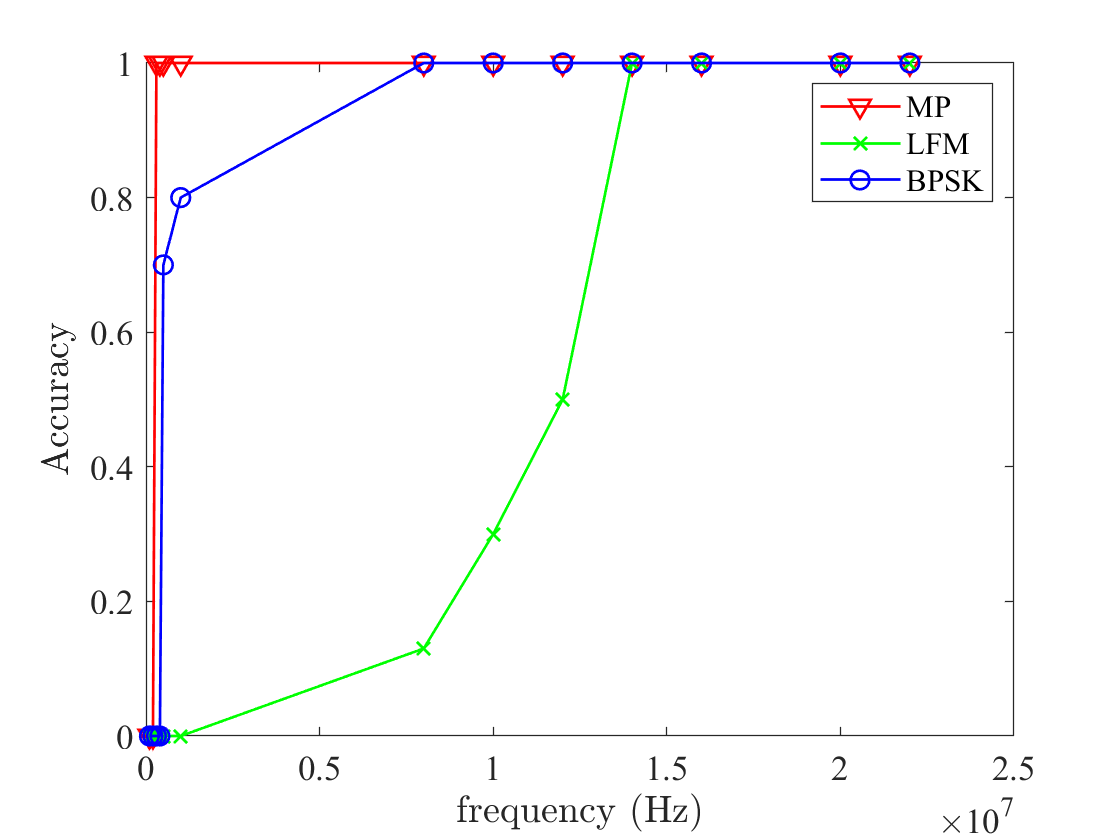}
	\caption{Power spectrum sensing performance versus frequency of phase modulation signal (K = 1, SNR = -5dB).}
	\label{fsin}
\end{figure}

In view of the influence of frequency and amplitude of phase modulation signal, power spectrum sensing accuracy results are compared as a function of the frequency of phase modulation signal shown in Figure~\ref{fsin}, under an input SNR of -5dB. There is $K = 1$ carrier frequency for each signal modulation type and that is selected in the 2-18GHz randomly. The symbol rate of the BPSK signal is set to 10M symbols per second with random code. And the bandwidth of the LFM signal is set to 10MHz. Meanwhile, the pulse length is fixed at 500ns with a start time randomly distributed throughout the observation duration.

As a result, the power spectrum sensing accuracy increases with the frequency of the sinusoidal phase modulation signal, and its performance closely follows the bandwidth of the interested signal. This is because the larger bandwidth of inputs such as the LFM signal, the more information is required for power spectrum reconstruction. As well as, the larger frequency of the phase modulation signal, there is the more number of RF non-uniform sampling pulses and the more information obtained by sampling.

\begin{figure}[htp]
	\includegraphics[width=8.5cm]{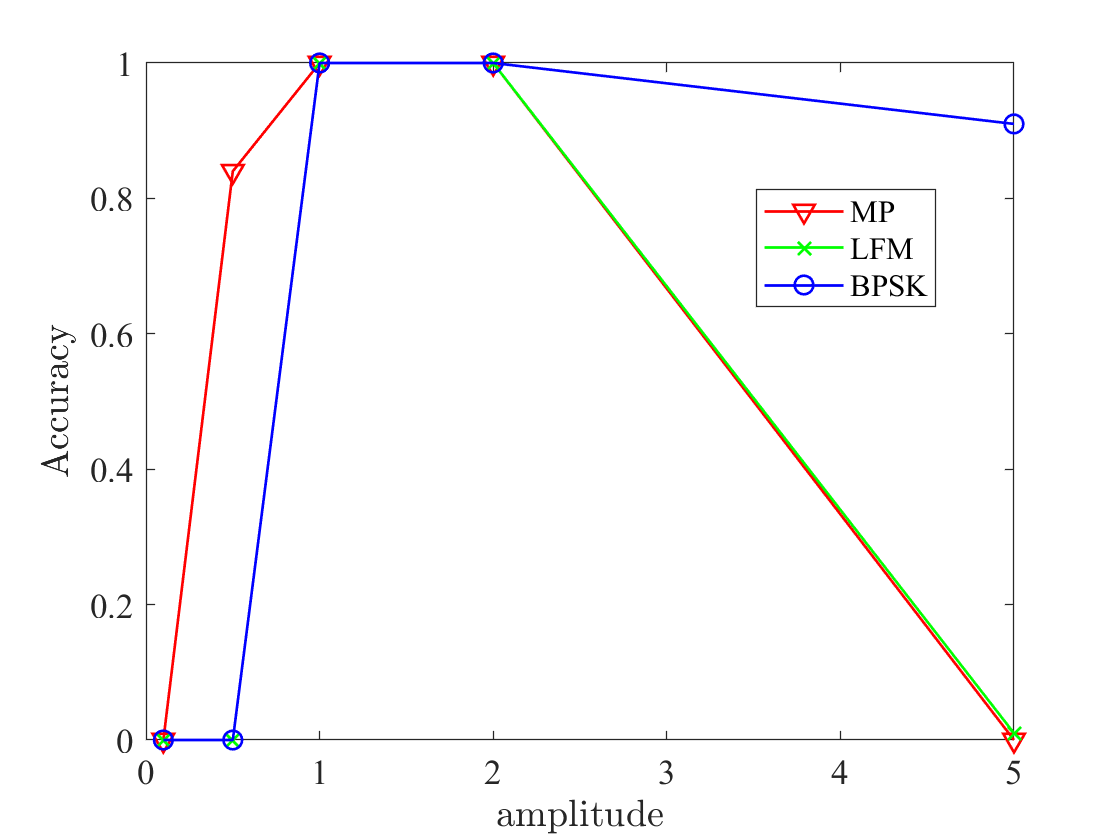}
	\caption{Power spectrum sensing performance versus amplitude of phase modulation signal (K = 1, SNR = -5dB).}
	\label{Asin}
\end{figure} 

As for the influence of the amplitude of the phase modulation signal, power spectrum sensing accuracy results are compared as a function of the amplitude of the phase modulation signal shown in Figure~\ref{Asin}, under an input SNR of -5dB. There is also set to $K = 1$ carrier frequency for each signal modulation type, which is selected in the 2-18GHz randomly. And then, the start time of the pulses is randomly distributed throughout the observation duration with the 500ns pulse length. Moreover, the symbol rate of the BPSK signal with random code and the bandwidth of the LFM signal are both set to 10MHz. 

As a result, the power spectrum sensing accuracy curve does not vary graphically as semilogarithmic with the amplitude of the sinusoidal phase modulation signal. There is the best power spectrum sensing performance for the three types of examples when the amplitude of the sinusoidal phase modulation signal is set to 1 and 2. This is because the larger amplitude of the phase modulation signal, there is the wider bandwidth of folded signal and the more dispersed signal energy, leading to the more serious aliasing of the sub-sampling harmonic. But for the smaller amplitude of phase modulation signal, the bandwidth expansion of the folded signal is not enough to distinguish between different NZ, especially for large bandwidth signals. Therefore, the optimization selection of system parameters is a multi-dimension, nonlinear, and complicated optimization problem, which is worthy of further research and exploration as a separate topic.

\begin{figure}[htp]
	\includegraphics[width=8.5cm]{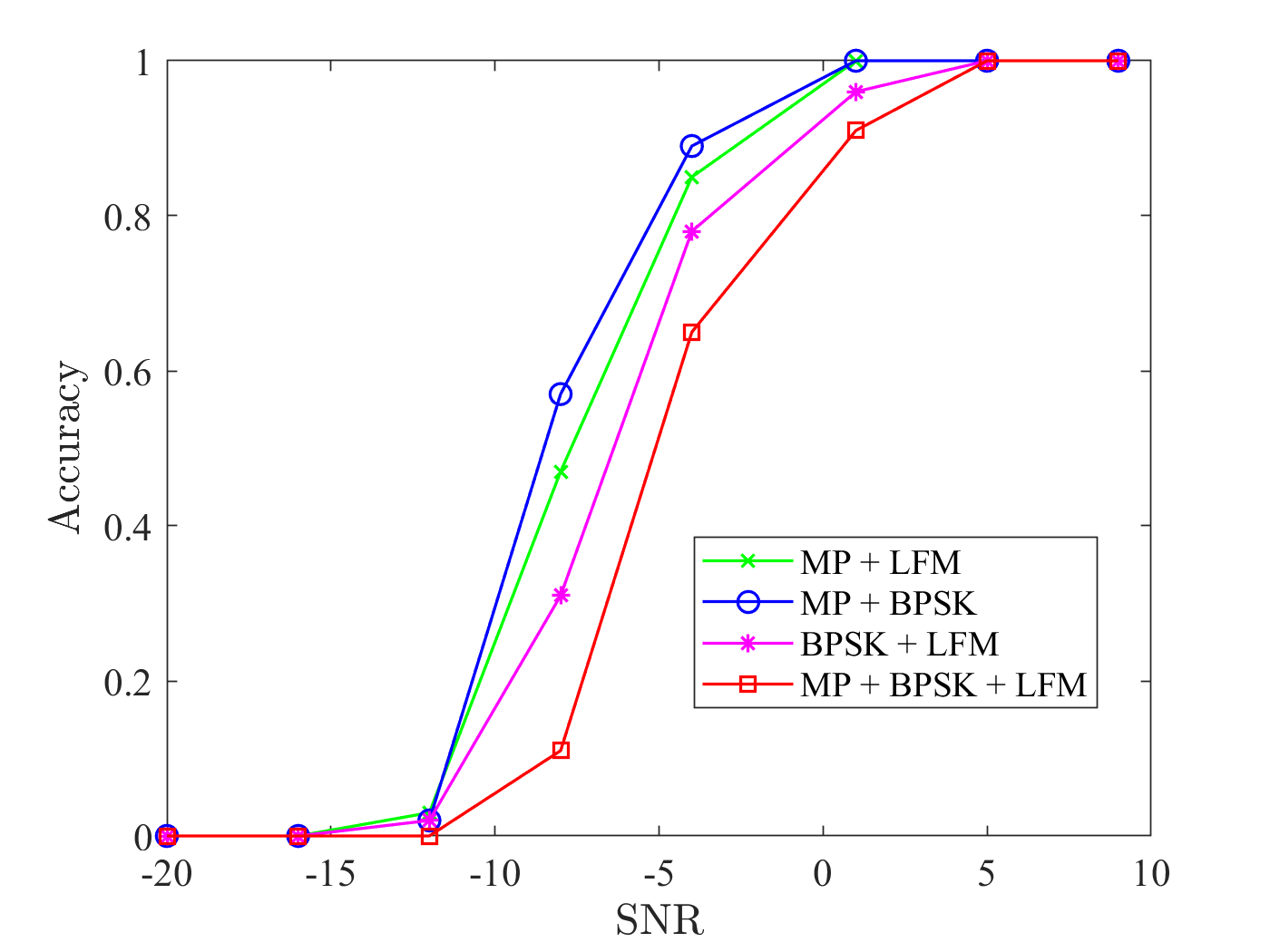}
	\caption{Power spectrum sensing performance versus SNR for multiple signals.}
	\label{multisig}
\end{figure}

Finally, the power spectrum sensing performance versus input SNR for multiple simultaneous received signals is displayed in Figure~\ref{multisig}. The simulation takes the combination of different signal modulation types as an example. The carrier frequency is selected in the 2-18GHz randomly for each signal modulation type. And the pulse length is no less than 500ns with the random start time of the pulses throughout the observation duration. Then, the symbol rate of the BPSK signal with random code and the bandwidth of the LFM signal are both set to 10MHz. Further, the amplitude of MP, BPSK, and LFM signals are respectively set to 1, 1.2, and 1.5. 

It is observed that the input SNR required for power spectrum sensing is increased with the increase of the number of simultaneous signal processing and signal modulation types. In addition, the best power spectrum sensing performance is for the combination of MP and BPSK signals, followed by the combination of MP and LFM signals, the combination of BPSK and LFM signals, and the combination of MP, BPSK, and LFM signals. It is evident that the power spectrum sensing performance is also limited by the bandwidth occupied by the received signal, which can be improved through the long time accumulation and smoothing.

In this following discussion, the computational complexity is contrasted of power spectrum sensing methods between the original time-domain algorithm, directly frequency-domain technique, and the proposed method. As for the direct original algorithms, the matrix $(\mathbf{A}^{*} \otimes \mathbf{A})\mathbf{C}$ and $\mathbf{B}^{*} \odot \mathbf{B}$ can be computed offline in advance according to (\ref{original}). Then, there are only three steps for the original time-domain algorithm, namely, calculating the correlation matrix $\mathbf{R}_y$, estimating the autocorrelation sequence $\mathbf{r}_s$, and taking the FT of $\mathbf{r}_s$. Thus, the time-domain algorithm requires $L M^2 + N M^2 + (2N-1) \text{log} (2N-1)$ floating-point operations in total, where $L$ is the smoothing factor of samples used to obtain the correlation matrix. Similarly, there are also three steps for the directly frequency-domain technique, namely, taking the FT of $\mathbf{y}(\omega_m)$, calculating the correlation matrix $\mathbf{R}_y (\omega)$, and computing a matrix-vector product in (\ref{original}). And the frequency-domain technique requires $M \text{log} M + L M^2 + N M^2$ floating-point operations in total. 

The proposed wideband power spectrum sensing method only involves NFT/FFT/IFFT operations and some multiplication calculations, which require computing (\ref{eq4}), (\ref{eq22}), (\ref{eq23}) and (\ref{eq24}). The autocorrelation sequence of $\mathbf{r}_p$ can be computed offline in advance according to (\ref{eq15}), which is only dependent on the RF non-uniform sampling pattern of the NYFR. The NFT of an N-point sequence for a k-sparse signal is first taken according to (\ref{eq4}), which involves $kN \text{log} N$ floating-point operations. And the IFFT of an N-point sequence is performed once with $(N \text{log} N)$ floating-point operations according to (\ref{eq22}). Additionally, the FFT of a $(2N-1)$-point sequence is done three times according to (\ref{eq23}) and (\ref{eq24}), which involves $(N \text{log} N) + (2N-1) \text{log} (2N-1)$ floating-point operations. Finally, $2N-1$ multiplication calculations are added. Therefore, the total computational complexity is $(k+1) N \text{log} N + (6N-3) \text{log} (2N-1) + (2N-1)$ floating-point operations, which scales linearly with the Nyquist-rate sampled number of samples N and the sparsity of spectrum occupancy k. 

Following the simulation parameters in section \ref{sec4}, to sense a wide spectrum range of 2-18GHz with the spectrum resolution of 500kHz, the direct original methods involve at least $5 \times 10^11$ floating-point operations in total with $L=100$. While for the proposed method, there are only $8.4 \times 10^6$ floating-point operations in total with $k=10$. Furthermore, if the spectrum occupancy reaches $70\%$, the total number of floating-point operations is as large as $10^{10}$ by using the proposed method for NYFR. Therefore, the proposed wideband power sensing method for NYFR has a lower computational complexity than the state-of-the-art methods, which meets the more practical solution for real-time applications.

\section{Conclusions}
\label{sec5}
Compressive power spectrum reconstruction is a more competitive solution for wideband spectrum sensing, which is computationally efficient and does not need any sparsity requirement, in the low SNR environment. The power spectrum estimation problem of NYFR is discussed for the first time in the existing, which can effectively avoid the weak signals swamped in the widening bandwidth or folded noise. The NYFR architectures can achieve a hundred percent of probability of intercept for full-band spectrum sensing by only a low-speed ADC with the low-speed circuits. The existing CCS-based methods are not suitable for NYFR with the high complexity and complicated calculations. By exploring the sub-sampling principle inherent in NYFR, a computationally efficient method is introduced with only the non-uniform fast Fourier transform, fast Fourier transform, and some simple multiplication operations. And its computational complexity isscales linearly with the Nyquist-rate sampled number of samples and the sparsity of spectrum occupancy. Simulation results and discussion demonstrate that the proposed method is a more practical solution to meet the real-time wideband spectrum sensing applications, compared with the state-of-the-art power spectrum sensing method in time-domain, with the low complexity in sampling and computation. Furthermore, the system optimization design of NYFR as well as peak spurious spectral suppressed are worthy of further investigation.

\section*{Acknowledgments}
The authors would like to thank all editors and reviewers for their valuable comments and suggestions.

\bibliographystyle{IEEEbib}
\bibliography{Refs_Manuscript}

\begin{thebibliography}{10}

\bibitem{Laake_2022_Remote}
Andreas Laake,
\newblock {\em Remote Sensing for Hydrocarbon Exploration},
\newblock Springer Remote Sensing/Photogrammetry. Springer International
  Publishing, 2022.

\bibitem{Arjoune_2019_Comprehensive}
Youness Arjoune and Naima Kaabouch,
\newblock ``A comprehensive survey on spectrum sensing in cognitive radio
  networks: Recent advances, new challenges, and future research directions,''
\newblock {\em Sensors}, vol. 19, no. 1, pp. 126, 2019.

\bibitem{Fang_2021_Recent}
Jun Fang, Bin Wang, Hongbin Li, and Ying-Chang Liang,
\newblock ``Recent advances on sub-nyquist sampling-based wideband spectrum
  sensing,''
\newblock {\em IEEE Wireless Commun.}, vol. 28, no. 3, pp. 115--121, 2021.

\bibitem{Manz_2021_Technology}
Barry Manz,
\newblock ``Technology survey: A sampling of analog-to-digital converter (adc)
  boards,''
\newblock {\em J. Electromagn. Domin.}, vol. 44, no. 2, pp. 25--30, 2021.

\bibitem{Pace_2022_Developing}
Phillip~E. Pace,
\newblock {\em Developing Digital RF Memories and Transceiver Technologies for
  Electromagnetic Warfare},
\newblock Artech House Electronic Warfare Library. Artech House, first edition
  edition, 2022.

\bibitem{Tsui_2015_Digital}
James Tsui and Chi-Hao Cheng,
\newblock {\em Digital Techniques for Wideband Receivers},
\newblock Radar, Sonar and Navigation. Institution of Engineering and
  Technology, third edition edition, 2015.

\bibitem{Mishra_2017_Compressivea}
Amit~Kumar Mishra and Ryno~Strauss Verster,
\newblock {\em Compressive Sensing Based Algorithms for Electronic Defence},
\newblock Signals and Communication Technology. Springer International
  Publishing, 2017.

\bibitem{Eldar_2012_Compresseda}
Yonina~C. Eldar and Gitta Kutyniok, Eds.,
\newblock {\em Compressed Sensing: Theory and Applications},
\newblock Cambridge University Press, 2012.

\bibitem{Zhou_2023_Structured}
Chengwei Zhou, Yujie Gu, Zhiguo Shi, and Martin Haardt,
\newblock ``Structured nyquist correlation reconstruction for doa estimation
  with sparse arrays,''
\newblock {\em IEEE Trans. Signal Process.}, vol. 71, pp. 1849--1862, 2023.

\bibitem{Liu_2022_SparsityBased}
Donghe Liu, Yongbo Zhao, and Tingxiao Zhang,
\newblock ``Sparsity-based two-dimensional doa estimation for co-prime planar
  array via enhanced matrix completion,''
\newblock {\em Remote Sensing}, vol. 14, no. 19, pp. 4690, 2022.

\bibitem{Huang_2022_Jointa}
Xiangdong Huang, Xuecheng Zhao, and Jinying Ma,
\newblock ``Joint carrier and doa estimation for multi-band sources based on
  sub-nyquist sampling coprime array with large time lags,''
\newblock {\em Signal Processing}, vol. 195, pp. 108466, 2022.

\bibitem{Joshi_2020_Learning}
Himani Joshi, Mohammad Alaee-Kerahroodi, A~Anil Kumar, Bhavani Shankar
  Mysore~R, and Sumit~J. Darak,
\newblock ``Learning based reconfigurable sub-nyquist sampling framework for
  ultra-wideband angular sensing,''
\newblock in {\em ICASSP 2020 - 2020 IEEE International Conference on
  Acoustics, Speech and Signal Processing (ICASSP)}, 2020, pp. 4637--4641.

\bibitem{Guo_2022_Dualband}
Liping Guo, Chi~Wah Kok, Hing~Cheung So, and Wing~Shan Tam,
\newblock ``Dual-band signal reconstruction based on periodic nonuniform
  sampling at optimal sampling rate,''
\newblock {\em Digital Signal Processing}, vol. 120, pp. 103252, 2022.

\bibitem{Liu_2022_novel}
Sujuan Liu, Lin Zhao, and Shibo Li,
\newblock ``A novel all-digital calibration method for timing mismatch in
  time-interleaved adc based on modulation matrix,''
\newblock {\em IEEE Trans. Circuits Syst. I}, vol. 69, no. 7, pp. 2955--2967,
  2022.

\bibitem{Jiang_2022_Joint}
Siyi Jiang, Ning Fu, Zhiliang Wei, Xiaodong Li, Liyan Qiao, and Xiyuan Peng,
\newblock ``Joint spectrum, carrier, and doa estimation with beamforming mwc
  sampling system,''
\newblock {\em IEEE Trans. Instrum. Meas.}, vol. 71, pp. 1--15, 2022.

\bibitem{Li_2021_Wideband}
Qiuyue Li, Zhi Li, and Jian Li,
\newblock ``Wideband spectrum sensing based on modulated wideband converter
  with nested array,''
\newblock {\em IET Communications}, vol. 15, no. 2, pp. 224--231, 2021.

\bibitem{Byambadorj_2020_Theoretical}
Zolboo Byambadorj, Koji Asami, Takahiro~J. Yamaguchi, Akio Higo, Masahiro
  Fujita, and Tetsuya Iizuka,
\newblock ``Theoretical analysis of noise figure for modulated wideband
  converter,''
\newblock {\em IEEE Trans. Circuits Syst. I}, vol. 67, no. 1, pp. 298--308,
  2020.

\bibitem{L3HARRIS__NYFR}
L3HARRIS,
\newblock ``Nyfr elint system,''
  \url{https://www.l3harris.com/all-capabilities/nyfr-elint-system}.

\bibitem{Fudge_2022_Multiple}
Gerald~L. Fudge, Ryan Lange, Calvin~A. Coffey, Frank~A. Boyle, Cameron Johnson,
  and Christopher~A. Fox,
\newblock ``Multiple clock sampling for nyquist folded sampling
  receivers.pdf,'' 2022.

\bibitem{Maleh_2012_Analogtoinformation}
Ray Maleh, Gerald~L. Fudge, Frank~A. Boyle, and Phillip~E. Pace,
\newblock ``Analog-to-information and the nyquist folding receiver,''
\newblock {\em IEEE J. Emerg. Sel. Topics Circuits Syst.}, vol. 2, no. 3, pp.
  564--578, 2012.

\bibitem{Wan_2023_Deepa}
Tao Wan, Kai-li Jiang, Hao Ji, and Bin Tang,
\newblock ``Deep learning-based lpi radar signals analysis and identification
  using a nyquist folding receiver architecture,''
\newblock {\em Defence Technology}, vol. 19, pp. 196--209, 2023.

\bibitem{Li_2018_Parameterc}
Tao Li, Qian Zhu, and Zengping Chen,
\newblock ``Parameter estimation of sar signal based on svd for the nyquist
  folding receiver,''
\newblock {\em Sensors}, vol. 18, no. 6, pp. 1768, 2018.

\bibitem{Tian_2022_Widebanda}
Kai-lun Tian, Kai-li Jiang, Sen Cao, Jian Gao, Ying Xiong, Bin Tang, Xu-ying
  Zhang, and Yan-fei Li,
\newblock ``Wideband spectrum sensing using step-sampling based on the
  multi-path nyquist folding receiver,''
\newblock {\em Defence Technology}, p. S2214914722002860, 2022.

\bibitem{Chae_2023_Rethinking}
Keunhong Chae, Jungin Park, and Yusung Kim,
\newblock ``Rethinking autocorrelation for deep spectrum sensing in cognitive
  radio networks,''
\newblock {\em IEEE Internet Things J.}, vol. 10, no. 1, pp. 31--41, 2023.

\bibitem{Wang_2018_Phasedarraybased}
Feiyu Wang, Jun Fang, Huiping Duan, and Hongbin Li,
\newblock ``Phased-array-based sub-nyquist sampling for joint wideband spectrum
  sensing and direction-of-arrival estimation,''
\newblock {\em IEEE Trans. Signal Process.}, vol. 66, no. 23, pp. 6110--6123,
  2018.

\bibitem{Ariananda_2012_Compressive}
Dyonisius~Dony Ariananda and Geert Leus,
\newblock ``Compressive wideband power spectrum estimation,''
\newblock {\em IEEE Trans. Signal Process.}, vol. 60, no. 9, pp. 4775--4789,
  2012.

\bibitem{Zhang_2021_Joint}
Zhan Zhang, Ping Wei, Huaguo Zhang, and Lijuan Deng,
\newblock ``Joint spectrum sensing and doa estimation with sub-nyquist
  sampling,''
\newblock {\em Signal Processing}, vol. 189, pp. 108260, 2021.

\bibitem{Xu_2019_efficient}
Wenbo Xu, Shu Wang, Shu Yan, and Jianhua He,
\newblock ``An efficient wideband spectrum sensing algorithm for unmanned
  aerial vehicle communication networks,''
\newblock {\em IEEE Internet Things J.}, vol. 6, no. 2, pp. 1768--1780, 2019.

\bibitem{Cohen_2014_SubNyquist}
Deborah Cohen and Yonina~C. Eldar,
\newblock ``Sub-nyquist sampling for power spectrum sensing in cognitive
  radios: A unified approach,''
\newblock {\em IEEE Trans. Signal Process.}, vol. 62, no. 15, pp. 3897--3910,
  2014.

\bibitem{Yang_2020_Fasta}
Linxiao Yang, Jun Fang, Huiping Duan, and Hongbin Li,
\newblock ``Fast compressed power spectrum estimation: Toward a practical
  solution for wideband spectrum sensing,''
\newblock {\em IEEE Trans. Wireless Commun.}, vol. 19, no. 1, pp. 520--532,
  2020.

\bibitem{Zhang_2018_Distributed}
Xingjian Zhang, Yuan Ma, Haoran Qi, Yue Gao, Zhixun Xie, Zhiqin Xie, Minxiu
  Zhang, Xiaodong Wang, Guangliang Wei, and Zheng Li,
\newblock ``Distributed compressive sensing augmented wideband spectrum sharing
  for cognitive iot,''
\newblock {\em IEEE Internet Things J.}, vol. 5, no. 4, pp. 3234--3245, 2018.

\bibitem{Ariananda_2011_Multicoset}
Dyonisius~Dony Ariananda, Geert Leus, and Zhi Tian,
\newblock ``Multi-coset sampling for power spectrum blind sensing,''
\newblock in {\em 2011 17th International Conference on Digital Signal
  Processing (DSP)}, 2011, pp. 1--8.

\bibitem{Wei_2022_Nonuniform}
Deyun Wei and Jun Yang,
\newblock ``Non-uniform sparse fourier transform and its applications,''
\newblock {\em IEEE Trans. Signal Process.}, vol. 70, pp. 4468--4482, 2022.

\end{thebibliography}

\vfill

\end{document}